\documentclass{article}%
\usepackage{amsmath}%
\usepackage{amsfonts}%
\usepackage{amssymb}%
\usepackage{graphicx}

\begin{document}
\title{A Study of the Dirac-Sidharth Equation}

\author{Raoelina Andriambololona\\
\textit{\small{Madagascar-Institut des Sciences et Techniques Nucl\'{e}aires, Madagascar-INSTN}}\\
\small{e-mail: raoelinasp@yahoo.fr}\\
Christian Rakotonirina
\\\textit{\small{Institut of High Energy Physics-Madagascar, iHEP-MAD}}\\
 \textit{\small{Institut Sup\'{e}rieur de Technologie d'Antananarivo, IST-T, Madagascar}}
\\\small{e-mail: rakotopierre@refer.mg}}

\maketitle

\begin{abstract}The Dirac-Siddharth Equation has been constructed from the Sid-
dharth hamiltonian by quantization of the energy and momentum in Pauli
algebra. We have solved this equation by using tensor product of matrices.
\end{abstract}

\section{Introduction}
In the special relativity of Einstein (A.Einstein., 1905), from the energy-momentum relation
\begin{equation}
E^2=c^{2}p^{2}+m^{2}c^{4}
\end{equation}
we can deduce the Klein-Gordon equation and the Dirac equation. This theory use the concept of continuous spacetime.  \\

Quantized spacetime was introduced at the first time by Snyder (H.S. Snyder., Phys Rev. 1947)\cite{Snyder472,Snyder471}, which known as Snyder noncommutative geometry. That is because the commutation relations are modified and becomes \cite{Snyder472,Snyder471}\\
\begin{equation}
\left[x^{\mu},x^{\nu}\right]=i\alpha\frac{\ell^{2}c^{2}}{\hbar}\left(x^{\mu}p^{\nu}-x^{\nu}p^{\mu}\right),
\end{equation}
\begin{equation}
\left[x^{\mu},p_{\nu}\right]=i\hbar\left[\delta^{\mu}_\nu+i\alpha\frac{\ell^{2}c^{2}}{\hbar^{2}}p^{\mu}p_{\nu}\right],
\end{equation}
\begin{equation}
\left[p_{\mu},p_{\nu}\right]=0
\end{equation}

\begin{equation}
\epsilon=\frac{\hbar c}{\sqrt{\alpha}\ell}
\end{equation}
is the energy due to the length scale $\ell$, where $\alpha$ a dimensionless constant.
As consequence the energy momentum relation gets modified and becomes \cite{Sidharth04}
\begin{equation}
E^2=c^{2}p^{2}+m^{2}c^{4}+\alpha\left(\frac{c}{\hbar}\right)^{2}\ell^{2}p^{4}
\end{equation}
$\ell=\ell_p=\sqrt{\frac{\hbar G}{c^{3}}}\approx1.6\times10^{-33}cm$, Planck scale, the fundamental length scale, where G is the gravitational constant.\\

 $\ell_c=\frac{e^{2}}{m_ec^{2}}\propto10^{-12}cm$, Compton scale, where $e$ is the electron charge and $m_e$ the electron mass.\\

$\ell_{LHC}\approx2\times10^{-18}cm$. \\
\noindent So,
\begin{equation}
\ell_p < \ell_{LHC} < \ell_c
\end{equation}
From the above energy- momentum relation, Sidharth has deduced the so-called Dirac-Sidharth Equation \cite{Sidharth04,Sidharth05}, a modified Dirac equation.\\
In the Section.2, we will derive the Dirac-Sidharth Equation by quantizing energy and momentum. In the Section.3, we will solve the Dirac-Sidharth Equation by using tensor product of matrices. \\

We think that using differents mathematical tools in physics will make to appear differents hidden mathematical or physical properties.\\

\section{A derivation of the Dirac-Sidharth equation}
For deriving the Dirac-Sidharth equation we use the method used by J.J. Sakurai \cite{Sakurai67} for deriving the Dirac equation.

The wave function of a spin-$\frac{1}{2}$ particle is two components. So, for quantizing the energy-momentum relation in order to have the modified Klein-Gordon equation \cite{Sidharth04,Sidharth05}, or Klein-Gordon-Sidharth equation, of the spin-$\frac{1}{2}$ particle, the operators which take part in the quantization should be $2\times2$ matrices. So, let us take as quantization rules

$E   \longrightarrow  i\hbar\sigma^{0}\frac{\partial}{\partial t}=i\hbar\frac{\partial}{\partial t}$

$\vec{p} \longrightarrow  -i\hbar\sigma^{1}\frac{\partial}{\partial x^{1}}-i\hbar\sigma^{2}\frac{\partial}{\partial x^{2}}-i\hbar\sigma^{3}\frac{\partial}{\partial x^{3}}=-i\hbar\vec{\sigma}\vec{\nabla}=\hat{p}_{1}\sigma^{1}+\hat{p}_{2}\sigma^{2}+\hat{p}_{3}\sigma^{3}$

\noindent where $\sigma^{1}$, $\sigma^{2}$, $\sigma^{3}$ are the Pauli matrices.
Then we have, at first the Klein-Gordon-Sidharth equation
\begin{equation}
c^{2}\hbar^{2}\left(\frac{\partial^{2}}{c^{2}\partial t^{2}}-\Delta-m^{2}c^{2}-\alpha\frac{\ell^{2}}{\hbar^{2}}\vec{\nabla}^{4}\right)\phi=0
\end{equation}
\begin{equation}
\begin{split}
\left(i\hbar\frac{\partial}{\partial t}+ic\hbar\vec{\sigma}\vec{\nabla}\right)\frac{1}{mc^{2}}\left\{\sum^{+\infty}_{k=0}\left(-1\right)^k\left [\frac{i\sqrt{\alpha}}{mc\hbar}\ell \left(-i\hbar\vec{\sigma}\vec{\nabla}\right)^{2}\right]^k\right\}\times \\ \left(i\hbar\frac{\partial}{\partial t}-ic\hbar\vec{\sigma}\vec{\nabla}\right)\phi
 = \left[mc^{2}+i\sqrt{\alpha}\frac{c}{\hbar}\ell \left(-i\hbar\vec{\sigma}\vec{\nabla}\right)^{2}\right]\phi
 \end{split}
\end{equation}
\noindent with application of the operator to two components wave function $\phi$, which is solution of the Klein-Gordon-Sidharth equation.
Let
\begin{equation}
\chi=\frac{1}{mc^{2}}\left\{\sum^{+\infty}_{k=0}\left(-1\right)^k\left [\frac{i\sqrt{\alpha}}{mc\hbar}\ell \left(-i\hbar\vec{\sigma}\vec{\nabla}\right)^{2}\right]^k\right\}\times \\ \left(i\hbar\frac{\partial}{\partial t}-ic\hbar\vec{\sigma}\vec{\nabla}\right)\phi
\end{equation}
\noindent then, we have the following system of partial differential equations
\begin{equation}
\left\{\begin{aligned}
i\hbar\frac{\partial}{c\partial t}\chi+i\hbar\vec{\sigma}\vec{\nabla}\chi & = mc\phi+i\sqrt{\alpha}\frac{\ell}{\hbar} \left(i\hbar\vec{\sigma}\vec{\nabla}\right)^{2}\phi\\
i\hbar\frac{\partial}{c\partial t}\phi-i\hbar\vec{\sigma}\vec{\nabla}\phi & = mc\chi-i\sqrt{\alpha}\frac{\ell}{\hbar} \left(i\hbar\vec{\sigma}\vec{\nabla}\right)^{2}\chi
\end{aligned} \right.
\end{equation}
In additionning and in subtracting these equations, and in transforming the obtained equation under matricial form, we have the Dirac-Sidharth equation
\begin{equation}
i\hbar\gamma^{\mu}_{D}\partial_{\mu}\psi_{D}-mc\psi_{D}-i\sqrt{\alpha}\ell\hbar\gamma^{5}_{D}\Delta\psi_{D}=0
\end{equation}
\noindent in the Dirac (or "Standard") representation of the $\gamma$-matrices, where \\
$\gamma^{0}_{D}=\begin{pmatrix}
\sigma^{0} & 0 \\
0 & -\sigma^{0}
\end{pmatrix}=\sigma^{3}\otimes\sigma^{0}$, $\gamma^{j}_{D}=\begin{pmatrix}
0 & \sigma^{j} \\
-\sigma^{j} & 0
\end{pmatrix}=i\sigma^{2}\otimes\sigma^{j}$, $j=1, 2, 3$, \\
 $\gamma^{5}_{D}=i\gamma^{0}_{D}\gamma^{1}_{D}\gamma^{2}_{D}\gamma^{3}_{D}=\begin{pmatrix}
0 & \sigma^{0} \\
\sigma^{0} & 0
\end{pmatrix}=\sigma^{1}\otimes\sigma^{0}$, and $\psi_{D}=\begin{pmatrix}
\chi+\phi\\
\chi-\phi
\end{pmatrix}$,\\
 $\Delta=\frac{\partial^{2}}{\partial x^{2}_{1}}+\frac{\partial^{2}}{\partial x^{2}_{2}}+\frac{\partial^{2}}{\partial x^{2}_{3}}$.\\
We know that (J.D. Bjorken and S.D. Drell., 1964)\cite{BjorkenDrell64}
\begin{equation}
P\gamma^{5}=-\gamma^{5}P
\end{equation}
It follows that the Dirac-Sidharth equation is not invariant under reflections (B.G. Sidharth., Mass of the Neutrinos, 2009).
The equation
\begin{equation}
i\hbar\gamma^{\mu}_{W}\partial_{\mu}\psi_{W}-mc\psi_{W}-i\sqrt{\alpha}\ell\hbar\gamma^{5}_{W}\Delta\psi_{W}=0
\end{equation}
is the Dirac-Sidharth equation in the Weyl(or "chiral") representation, where  $\psi_{W}=\begin{pmatrix}
\chi\\
\phi
\end{pmatrix}$.\\
So, $\chi$ is the left-handed two components spinor and $\phi$ the right-handed one.
This method makes to appear that the right-handed two components spinor is solution of the Klein-Gordon-Sidharth equation.
\section{Resolution of the Dirac-Sidharth equation}
In this section we will use the tensor product of matrices for solving the Dirac-Sidharth equation. We had used this method,  suggested by Raoelina Andriambololona for solving the Dirac equation  (C. Rakotonirina., Thesis, 2003).\\
Let us look for a solution of the form
\begin{equation}
\psi_{D}=U(p)e^{\frac{i}{\hbar}\left(\vec{p}\vec{x}-Et\right)}
\end{equation}
Let $\Psi$ a four components spinor which is eigenstate both of $\hat{p}_{j}=-i\hbar\frac{\partial}{\partial x^{j}}$ and $\hat{E}=i\hbar\frac{\partial}{\partial t}$, $\vec{p}=\begin{pmatrix}
p^{1}\\
p^{2}\\
p^{3}
\end{pmatrix}$, and $\vec{n}=\frac{\vec{p}}{p}=\begin{pmatrix}
n^{1}\\
n^{2}\\
n^{3}
\end{pmatrix}$.\\
The Dirac-Sidharth equation becomes
\begin{equation}
\sigma^{0}\otimes\sigma^{0}U(p)-\frac{2}{\hbar}cp\sigma^{1}\otimes\left(\frac{\hbar}{2}\vec{\sigma}\vec{n}\right)U(p)-mc^{2}\sigma^{3}\otimes\sigma^{0}U(p)+c\sqrt{\alpha}p^{2}\frac{\ell}{\hbar}\sigma^{2}\otimes\sigma^{0}U(p)=0
\end{equation}
Let us take $U(p)$ of the form
\begin{equation}
U(p)=\varphi\otimes u
\end{equation}
where $u$ is the eigeinvector of the spin operator $\left(\frac{\hbar}{2}\vec{\sigma}\vec{n}\right)$. $\varphi=\begin{pmatrix}
\varphi^{1}\\
\varphi^{2}
\end{pmatrix}$ and $u$ are two components.\\
Since $u\neq0$, so
\begin{equation}
\left(\epsilon cp\sigma^{1}+mc^{2}\sigma^{3}-c\sqrt{\alpha}p^{2}\frac{\ell}{\hbar}\sigma^{2}\right)\varphi= E\varphi
\end{equation}
with $\epsilon=
\begin{cases}
+1 & \text{spin up}\\
-1 & \text{spin down}
\end{cases}$ \\
Solving this equation with respect to $\varphi^{1}$ and $\varphi^{2}$, we have
 \begin{equation}
\Psi_{+}=\sqrt{\frac{E+mc^{2}}{2E}}\begin{pmatrix}
1\\
\frac{\epsilon cp-i\frac{c}{\hbar}\sqrt{\alpha}p^{2}\ell}{mc^{2}+E}
\end{pmatrix}\otimes se^{\frac{i}{\hbar}\left(\vec{p}\vec{x}-Et\right)}
\end{equation}
the solution with positive energy, where $s=\frac{1}{\sqrt{2\left(1+n^{3}\right)}}\begin{pmatrix}
-n^{1}+in^{2}\\
1+n^{3}
\end{pmatrix}$ spin up, $s=\frac{1}{\sqrt{2\left(1+n_{3}\right)}}\begin{pmatrix}
1+n^{3}\\
n^{1}+in^{2}
\end{pmatrix}$ spin down.\\

This method makes to appear the $2\times2$ matrix $h=\epsilon cp\sigma^{1}-c\sqrt{\alpha}p^{2}\frac{\ell}{\hbar}\sigma^{2}+mc^{2}\sigma^{3}$ whose eigeinvalues are the positive and the negative energies. $h$ is like a vector in Pauli algebra. So, energy of the spin-$\frac{1}{2}$ particle can be associated to a vector in Pauli algebra, whose length or intensity is given by the energy-momentum relation.
\begin{equation}
h^{2}=E^{2}
\end{equation}


\begin{thebibliography}{9}                                                                                                %
\bibitem {Sidharth06}Sidharth, B.G., arXiv:physics/0603189
\bibitem {Snyder472} Snyder, H.S., Phys.Rev., Vol.72, No.1, July 1, 1947, 68-71.
\bibitem {Snyder471} Snyder, H.S., Phys.Rev., Vol.71, No.1, January 1, 1947, 38-41.
 \bibitem{Sidharth04}Sidharth, B.G., Int.J.Th.Phys., Vol.43, No.9, September 2004, 1857-1861.
\bibitem {Sidharth05}Sidharth, B.G., Int.J.Mod.Phys.E., Vol.14, No.6, 2005, 927-929.
\bibitem {Sakurai67}Sakurai, J.J., \textit{Advanced Quantum Mechanics}, Addison Wesley Publishing Company, 1967, 308-311.
\bibitem {BjorkenDrell64} J.D. Bjorken and S.D. Drell., \textit{Relativistic Quantum Mechanics}, Mc-Graw Hill, New York, 1964, pp.39.
\bibitem {Sidharth09}Sidharth, B.G., arXiv:0904.3639v1.
\end{thebibliography}
\end{document}